\documentclass[useAMS,usenatbib]{mn2e}

\usepackage{color}

\usepackage{aecompl}
\usepackage{graphicx}
\usepackage{amssymb}
\usepackage{amsmath}
\usepackage{multicol}
\usepackage{longtable}
\usepackage{algorithm}
\usepackage{longtable}
\usepackage{float}
\usepackage{here}

\usepackage[colorlinks=true]{hyperref}
\usepackage{lscape}

\title[Evolution of Brightest Cluster Galaxies]{ Stellar Mass Assembly of Brightest Cluster Galaxies at Late Times}

\author[T. Inagaki et al.]{\parbox[t]{\textwidth}{\vspace{-1cm}
Takahiro Inagaki$^1$, 
Yen-Ting Lin$^{2}$\thanks{E-mail:ytl@asiaa.sinica.edu.tw}, Hung-Jin Huang$^{2,3,4}$, Bau-Ching Hsieh$^{2}$, and Naoshi Sugiyama$^{1,5,6}$}\\
$^1$Department of Physics and Astrophysics, Nagoya University, Nagoya 464-8602, Japan\\
$^2$Institute of Astronomy and Astrophysics, Academia Sinica, Taipei 10617, Taiwan\\
$^3$Institute of Astrophysics, National Taiwan University, Taipei 10617, Taiwan\\
$^4$Department of Physics, Carnegie Mellon University, Pittsburgh, PA 15213, USA\\
$^5$Kavli Institute for the Physics and Mathematics of the Universe,
The University of Tokyo, Kashiwa, Chiba, 277-8568, Japan\\
$^6$Kobayashi-Maskawa Institute for the Origin of Particles and the Universe, Nagoya University, Nagoya 464-8602, Japan\\
}

\begin{document}

\newcommand{\arl}[1]{\url{#1}}
\newcommand{\eqnref}[1]{Eq.(\ref{#1}) }
\newcommand{\eqnrefs}[1]{Eq.(\ref{#1})}
\newcommand\mnras{MNRAS}
\newcommand\nat{Nature}
\newcommand{\aj}{Astronomical. J.}
\newcommand\apjs{ApJS}
\newcommand\apjl{ApJL}
\newcommand\aap{A\&A}
\newcommand\apss{Ap\&SS}
\newcommand\ssr{SPRINGER}
\newcommand\jgr{JGeophysRes}
\newcommand\icarus{Icarus}

\newcommand{\jcap}{JCAP}
\newcommand\apj{ApJ}
\newcommand\pasp{PASP}

\date{Accepted --. Received --; in original form --}

\pagerange{\pageref{firstpage}--\pageref{lastpage}} \pubyear{2014} 

\maketitle

\label{firstpage}

\begin{abstract}

Understanding the formation history of brightest cluster galaxies is an important topic in galaxy formation.
Utilizing the {\it Planck} Sunyaev-Zel'dovich cluster catalog,
and applying the Ansatz that the most massive halos at one redshift remain among the most massive ones at a slightly later cosmic epoch, we have constructed cluster samples at redshift $z\sim 0.4$ and $z\sim 0.2$ that can be statistically regarded  as progenitor-descendant pairs.  This allows us to study the stellar mass assembly history of BCGs in these massive clusters at late times, finding the degree of growth between the two epochs is likely at only few percent level, which is far lower compared to the prediction from a state-of-the-art semi-analytic galaxy formation model.

\end{abstract}


\section{Introduction}
\label{introduction}

In the cold dark matter dominated Universe, structure growth proceeds hierarchically through gravitation interactions \citep[e.g.,][]{white78,davis85,2005Natur.435..629S}. 
Galaxy clusters represent the culmination of structure formation at the present time, and most of the clusters are still in the active forming phase.  In particular, the high resolution $N$-body simulation of \citet{2004ApJ...614...17G} has shown that even at $z<1$, frequent mergers have brought lots of material into the very center of massive halos, namely where the brightest cluster galaxies (BCGs) in real clusters are located.  As such, it is natural to expect the BCGs to grow in mass at late times in cosmic history.  This is indeed a generic prediction of the semi-analytic models (SAMs); in particular, \citet{2007MNRAS.375....2D}  show that at $z<0.5$, their model BCGs typically have gained $\sim 50\%$ of their final stellar mass.

Some researchers have investigated the stellar mass of BCGs in massive clusters by using near-infrared luminosity as a mass proxy,  
and have suggested that BCGs grow little in mass since $z\sim 1$\citep{2009Natur.458..603C,2010ApJ...718...23S}.
Both \citet{2012MNRAS.427..550L} and \citet{2013ApJ...771...61L} have examined the stellar mass growth of BCGs across a wide range in redshifts, and have found a factor of $\sim 2$ growth between $z=1$ and 0.  In particular, \citet{2013ApJ...771...61L} use a sample of intermediate mass clusters and find that, although the BCG growth is consistent with the predictions from the latest version of Munich SAM built upon the Millennium Run simulation \citep{2005Natur.435..629S, 2011MNRAS.413..101G}, below $z=0.5$ there is some hint of divergent behavior between the model and observed BCGs.  While the model BCGs continue to grow in mass at about the same rate, the growth of observed BCGs seems to slow down considerably.  It is possible that for the real BCGs, the continuing accretion of satellite galaxies adds mass to the outskirts of the galaxies, far beyond the observed regions (typically $\sim 30$ kpc in diameter; \citealt{2008MNRAS.387.1253W}).  If true, this could offer an explanation of the discrepancy.

Clearly the epoch $z<0.5$ is an important period of time to examine the stellar mass assembly history of BCGs.  We would also like to extend the study of \citet{2013ApJ...771...61L} in terms of the mass range of clusters, to investigate the formation of BCGs in very massive clusters.  Thus, in the present study we will focus on the evolution of BCGs hosted by massive clusters at $z<0.5$.  
To achieve our goal, we need to make sure that within our cluster sample,
the higher redshift clusters are representative of the progenitors of the lower redshift clusters.
A novelty of our approach is the use of a fixed number density to select massive clusters, which we find can fulfill such a progenitor-descendant relationship requirement for the clusters. 
Furthermore, instead of studying the stellar mass within a fixed aperture, we attempt to constrain the ``total'' stellar mass of the BCGs, as allowed by the available data.

To this end, we have employed the cluster sample detected via the  Sunyaev-Zel'dvich (SZ) effect \citep{1970Ap&SS...7...20S, 1969Ap&SS...4..301Z} 
by the {\it Planck} satellite \citep{2013arXiv1303.5080P, 2013arXiv1303.5089P}.
The SZ effect is the inverse Compton scattering of Cosmic Microwave Background (CMB) photon by hot electrons in the intracluster medium.  
As the SZ flux is expected to correlate tightly with total thermal energy of the cluster, and thus its total mass \citep{2013arXiv1303.5080P},
selection via the SZ effect is expected to produce
a unbiased, massive cluster sample.  
We then apply the fixed number density selection to the {\it Planck} cluster sample, and utilize an Ansatz (see Section \ref{assumption}) that allows us to connect clusters at different redshifts as an evolutionary sequence to study the BCG evolution, using data from the Sloan Digital Sky Survey (SDSS; \citealt{2000AJ....120.1579Y}).

This paper is organized as follows. 
In Section \ref{assumption}, we describe the Ansatz, which forms the starting point of our analysis. 
In Section \ref{planck}, we present the details of our analysis of the {\it Planck} SZ cluster sample, including the identification of the BCGs and the estimation of their stellar mass. 
The results on the stellar mass assembly history of BCGs are shown in Section \ref{results_1}.
Finally we summarise our results in Section \ref{summary}.  
Unless otherwise noted, throughout this paper we use a simple $\Lambda$CDM cosmology model 
where $\Omega_m=0.3$, $\Omega_\Lambda=0.7$ and $H_0=100h~{\rm km~s^{-1}~Mpc^{-1}}$, with $h=0.7$.

\section{The Ansatz}
\label{assumption}

Observationally it is challenging to follow the evolution of any population of galaxies, although significant progresses have been made recently, such as the method that selects galaxies at or above a fixed (cumulative) number density \citep{2010ApJ...709.1018V, 2013ApJ...777...18M}.  However, there is a great advantage of working with galaxies in clusters, because once we could identify clusters that {\it statistically} form an evolutionary sequence, we can then link the galaxy populations in these clusters over cosmic time, and study their evolution\footnote{In practice one needs to take into account the fact that clusters continue to acquire galaxies via accretion and merger with surrounding galactic systems.}.  In this study we focus on the BCGs in clusters that form such an evolutionary sequence.

\begin{figure}
\includegraphics[width=.45\textwidth]{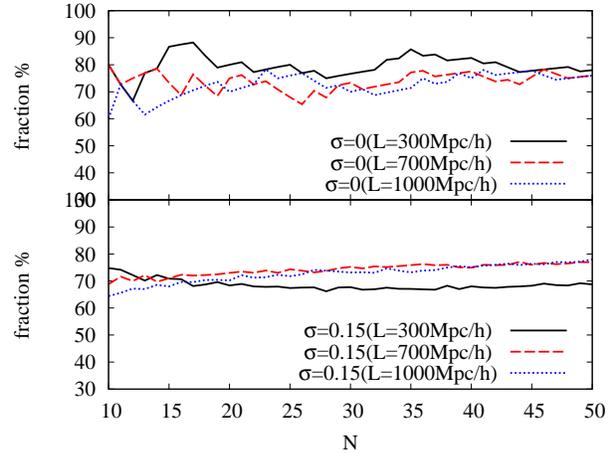}
\caption{
The fraction of the top $N$ halos at $z=0.4$ that remain among the top $N$ rank at $z=0.2$. 
The solid line, dashed line  and dotted line are for simulations with box sizes of 
$300$, $700$ and $1000 h^{-1}{\rm Mpc}$, respectively. 
Top panel: the ideal case, when there is no scatter between the cluster mass and the mass proxy used to select the clusters.  Bottom panel: the case when 15\% fractional scatter is applied before selecting the clusters, which should be applicable to the {\it Planck} clusters used in our analysis.
}
\label{fig:0}
\end{figure}

We construct such a cluster sample based on the {\it Ansatz} that, in a given comoving volume $V$, the top $N$ most massive clusters at one epoch will remain among the most massive ones at a slightly later epoch, separated by a period $\Delta t$.  We test what optimal $V$, $N$, and $\Delta t$ should be using cosmological $N$-body simulations, and consider the effect of scatter in the mass--observable relation, as follows.

As we are mainly interested in the BCG evolution at late times ($z<0.5$), we consider redshift ranges $0.13\le z \le0.26$ and $0.37 \le z \le 0.41$ (hereafter denoted as low-$z$ and high-$z$, respectively), which occupy the same comoving volume for the same solid angle.  At these redshifts, data from SDSS are adequate for studying luminous galaxies such as BCGs.

To check the Ansatz, we have performed three dark matter-only cosmological simulations with the massively parallelized code $Gadget$-2 in its Tree-PM mode 
\citep{2001NewA....6...79S,2005MNRAS.364.1105S}. 
The softening length is $3\%$ of the mean interparticle separation and the total particle number is $N_{tot}=1024^3$. 
The box sizes are $300$, $700$ and $1000~h^{-1}{\rm Mpc}$.  
The volume of the $700~h^{-1}{\rm Mpc}$ run is closest to the actual observations we have (see Section \ref{planck}).
For each simulation, we first select the top $N$ most massive halos at $z=0.4$, then use the merger history to identify their descendants at $z=0.2$.
We compare these descendants with the top $N$ most massive halos at $z=0.2$ and  
calculate the fraction of halos that are present in both halo samples.
Fig.~\ref{fig:0} (top panel) shows such fractions as a function of $N$. 
The simulations suggest that about 75\% of halos remain among the most massive ones between $z=0.2$ and $z=0.4$ with $N\approx 30$.
We also find that the effect of the box size is small for the cases where $N\ge 50$.

In practice, we usually cannot select clusters by their mass, but rather by a mass proxy, which inevitably exhibits some scatter with respect to the true mass.  As we will use the cluster sample detected by {\it Planck} via the SZ effect, we consider the 
effect of  scatter in the $M_{Y_Z}$--mass relation, where $M_{Y_Z}$ is the best mass proxy (with a scatter of $\sim 15\%$) recommended by the {\it Planck} team \citep{2013arXiv1303.5080P}.
In our simulation box at $z=0.4$, for each halo, we randomly perturb the halo mass by a random Gaussian variate with $\sigma=0.15$.  The same operation is applied to the halos at $z=0.2$ as well.  We then compute the fraction of descendants of the top $N$ $z=0.4$ halos selected by the perturbed mass that remain in the top $N$ list at $z=0.2$ (also selected by the perturbed mass).  The results are shown in Fig.~\ref{fig:0} (bottom panel).  
Interestingly, the introduction of scatter in mass proxies only has appreciable effects on the remaining fraction for our smallest simulation box, and we find that for larger boxes, which are closer in terms of the comoving volume to our actual observations, the effect is quite small, and our Ansatz holds to about 75\%.

\subsection{Inferring the stellar mass growth of BCGs}
\label{bcg_growth}

How does the scatter affect the robustness of our approach in inferring the BCG stellar mass growth?  We repeat the procedure just mentioned, but now, at $z=0.4$, for each halo we also assign a BCG stellar mass with a mean of $M_{bcg}=aM_h^c$ (where $a$ is an arbitrary normalisation factor and slope $c=0.2$) and a Gaussian scatter of $30\%$ at fixed (true/unperturbed) halo mass $M_h$. These values of $c$ and scatter are consistent with the measurement of \citet{lin04} for nearby X-ray clusters.  At $z=0.2$, a BCG stellar mass--halo mass relation with the same scatter and slope, but different normalisation, is used ($M_{bcg}=bM_h^c$).  
Now, there are three relevant quantities to consider: (1) the mean BCG stellar mass of top $N$ halos at $z=0.4$ selected by the perturbed mass, $\overline{M_{bcg,1}}$;   (2) the mean BCG stellar mass of top $N$ halos at $z=0.2$ selected by the perturbed mass, $\overline{M_{bcg,2}}$; (3) the mean BCG stellar mass at $z=0.2$ for descendants of the top $N$ $z=0.4$ halos selected by the perturbed mass, $\overline{M_{bcg,3}}$.  While observationally we measure $\overline{M_{bcg,1}}$ and $\overline{M_{bcg,2}}$, it is the ratio  $\overline{M_{bcg,3}}/\overline{M_{bcg,1}}$ that we are after.  
Under the fair assumption that the BCG stellar mass assembly in descendants  of top $N$ $z=0.4$ halos is the same as in other massive $z=0.2$ halos,
we could use this simple simulation to examine whether we could recover the overall growth of BCGs (i.e., $b/a$), as well as the growth of the particular BCG population we are interested in (i.e., the top $N$ halos at $z=0.4$ and their descendant halos).

\begin{figure}
\includegraphics[width=.45\textwidth]{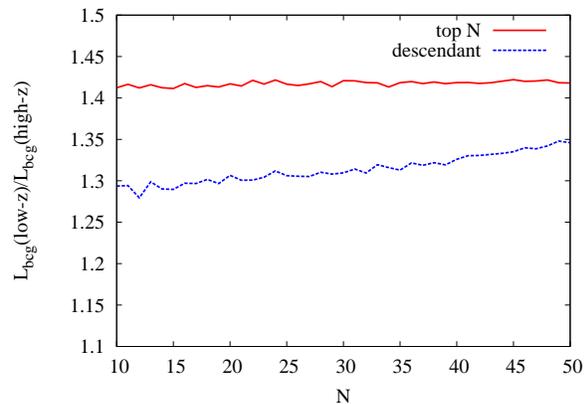}
\caption{
Recovery of the degree of BCG stellar mass growth, as a function of the top $N$ selection (using the $700 h^{-1}$\,Mpc box).  In this example, we have assumed that, between $z=0.4$ and $z=0.2$, the amplitude of the BCG stellar mass--cluster mass relation has increased by a factor of two, while the scatter in the relation remains at 30\%.   The red curve shows that, when the top $N$ clusters are selected at both redshifts, the ratio of the mean BCG stellar masses is biased high by $\sim 9\%$ compared to the actual amplitude.  The blue dotted curve shows the case when the descendants of top $N$ $z=0.4$ halos are compared against their progenitors.  In this case the resulting bias is quite small.
}
\label{fig:recovery}
\end{figure}

The results from this exercise are shown in Fig.~\ref{fig:recovery}, as a function of $N$.  We have arbitrarily set $b/a=1.3$, and found that while the ratio $\overline{M_{bcg,2}}/\overline{M_{bcg,1}}$ is biased high by 9\%, the ratio $\overline{M_{bcg,3}}/\overline{M_{bcg,1}}$ is very close to the actual value.  In principle, these results allow us to infer the true growth $\overline{M_{bcg,3}}/\overline{M_{bcg,1}}$ by applying a correction factor to the observed growth $\overline{M_{bcg,2}}/\overline{M_{bcg,1}}$.  In reality, however, since the exact magnitudes of scatter in both the cluster mass--observable relation and the BCG stellar mass--cluster mass relation are not well measured, later in the analysis we will mainly invoke the results here for qualitative arguments.

\section{The Analysis}
\label{planck}

In this section, we first describe the data sets used in our analysis.  These include
the {\it Planck} SZ catalog and the SDSS data used to study the BCGs.  For comparison with predictions from galaxy formation theories, we also make extensive use of the Millennium Run database.
We then provide detailed accounts of our identification of optical counterparts of {\it Planck} SZ sources and the designation of BCG (Section \ref{method_data}), and the estimation of stellar masses of BCGs (Section \ref{bcg_sm}).

{\it Planck} is a project of the European Space Agency \citep{2013arXiv1303.5062P}. 
The main goal of the mission is to determine the cosmological parameters describing the Universe. 
The 74 detectors of {\it Planck} satellite are sensitive to a range of frequencies from 
$30$ to $857$GHz. 
One of the most important results from {\it Planck} so far is the construction of an all-sky cluster catalog derived from the SZ effect,
using data from the first 15.5 months of observations 
\citep{2013arXiv1303.5089P}. 
The {\it Planck} Sunyaev-Zel'dovich (PSZ) catalogue contains 1227 clusters and is the largest SZ catalogue to date. 
We use this catalogue 
as our parent cluster sample.

The SDSS and its later incarnations have observed about one quarter of the sky \citep{2000AJ....120.1579Y}. 
The goal of the project is to create a 3-dimensional map of the Universe.  
It uses a 2.5m telescope equipped with a mosaic CCD camera which can image the sky in five broad optical bands, and a pair of multi-object spectrographs covering the whole optical wavelength.
We use the latest public data release from SDSS, DR10 \citep{2013arXiv1307.7735A}, to estimate the cluster redshifts, and to identify and study BCGs.

The Millennium Run  is a very high-resolution cosmological $N$-body simulation  
that follows the evolution of $N=2160^3$ particles from redshift $z=127$ to the present, with a box size of $500h^{-1}{\rm Mpc}$ 
\citep{2005Natur.435..629S}. 
Galaxy formation in Millennium is treated using semi-analytic prescriptions \citep[e.g.,][]{2011MNRAS.413..101G, 2006MNRAS.366..499D}. 
To compare with our observations, we use the predictions from one of these semi-analytic models (SAMs).

\subsection{Optical counterparts of PSZ clusters}
\label{method_data}

The first step in our analysis is to identify the galaxies associated with each of the PSZ sources.  This is to ensure the robustness of the SZ detections, and to validate the cluster redshifts from the PSZ catalogue.
We start with the 374 clusters that lie within the SDSS DR10 footprint, and have the ``validation'' flag value $\ge 10$ in the PSZ catalog, 
which indicates that these are either newly confirmed or previously-known clusters. 
Those with validation value $<10$ are discarded as they lack redshift information.

We use the PSZ catalogue to obtain the initial indication of the position of the clusters. 
The PSZ position uncertainty $\delta_x$, provided in the catalog for each cluster, can be up to a few arcmin.  We have thus queried the SDSS database within
$1 h^{-1}{\rm Mpc}+ \delta_x$ from the nominal cluster center
to obtain colour images and photometric and spectroscopic catalogs.

Although significant efforts have gone into the optical identification and verification of cluster candidates in the PSZ catalog, in some cases we cannot find unambiguously the optical counterpart of PSZ sources 
within the search radius, while in others the redshift listed in PSZ catalog does not agree with the apparent redshift of counterpart clusters we identify.
Therefore, in addition to the cluster redshift listed in the PSZ catalog ($z_p$), 
we determine the cluster redshift based on the mean redshift of red member galaxies in the cluster, as described below. 

\begin{figure}
\includegraphics[width=.45\textwidth]{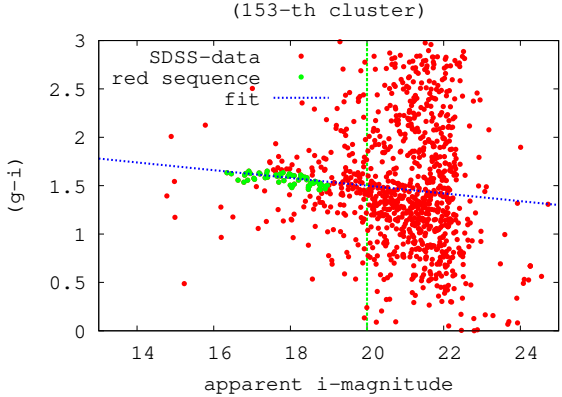}
\caption{$g-i$ vs $i$ colour-magnitude diagram for one of the clusters in our sample. 
The red and green points represent all galaxies and the red sequence galaxies, respectively. 
The blue line is the best-fit relation, obtained with a $3\sigma$ clipping algorithm.
}
\label{fig:1}
\end{figure}

For each of the PSZ candidate clusters with $z_p$ close to our two redshift bins, we visually inspect the SDSS image and look for the optical counterpart, in the form of spatial concentration of red galaxies.  
Operationally,
we first look for a red sequence in the $g-i$ vs $i$-band colour-magnitude space
(Fig. \ref{fig:1}). 
Old galaxies in a cluster are often found to lie on a narrow sequence defined by a colour index that straddles the $4000$\,\AA\ break \citep[e.g.,][]{2004ApJ...614..679L}.  
As our clusters span the redshift range from $z\sim 0.15$ to $z\sim 0.4$, 
for simplicity we choose the $g-i$ colour instead of the conventional $g-r$ to avoid the filter edge effect at around $z\sim 0.4$.

An example of the colour-magnitude diagram is shown in Fig. \ref{fig:1}.  The red sequence can be clearly seen.
The blue line is the best fit to the sequence, obtained from a $3\sigma$-clipping technique\footnote{In short, we first divide the colour-magnitude space into small cells and calculate the surface density of galaxies in the cells.  We then set a lower limit in surface density for regions to be used in fitting the red sequence.  This procedure effectively eliminates the blue cloud.  We then apply $3\sigma$-clipping to better measure the tilt and amplitude of the red sequence.}.
For the fit,  
we consider only galaxies brighter than an apparent magnitude limit $i_{\lim}$; for the low-$z$ bin we set $i_{lim}=20$, while for the high-$z$ bin $i_{lim}=22$.
To accommodate the finite width of the red sequence,
we assume the galaxies with colour in the range $-0.1< (g-i)-(g-i)_{\rm fit}<0.05$ belong to the  sequence.

\begin{figure}
\includegraphics[width=.45\textwidth]{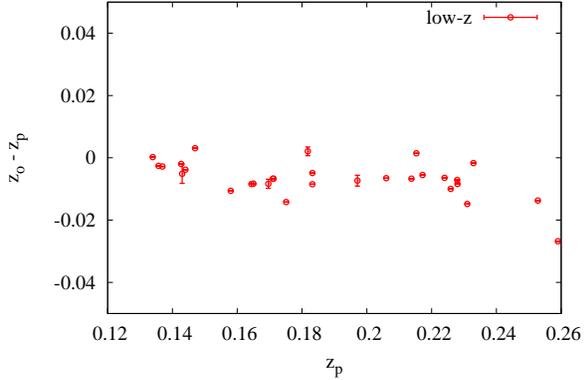}
\caption{
The difference between the cluster redshift determined by our method ($z_o$) and that provided by the {\it Planck} team ($z_p$) for the low-$z$ bin. $z_o$ and their uncertainties are obtained by weighted mean of photometric and spectroscopic redshifts of red sequence galaxies. Clusters with vanishingly small error bars are due to the presence of a few galaxies with consistent $z_{spec}$.   
}
\label{fig:4}
\end{figure}

\begin{figure}
\includegraphics[width=.45\textwidth]{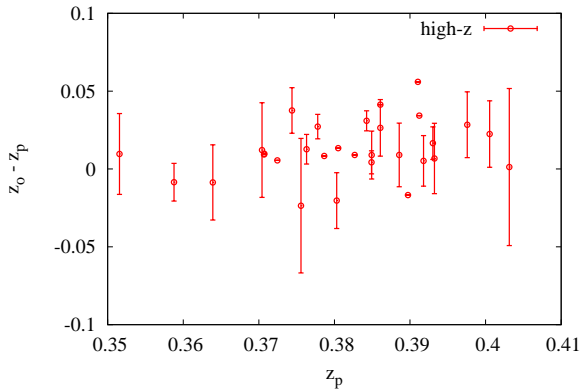}
\caption{Same as Fig.~\ref{fig:4}, but for the high-$z$ bin.}
\label{fig:5}
\end{figure}

In SDSS DR10, for bright galaxies that we consider ($i\le i_{lim}$), photometric redshift ($z_{phot}$) and in some cases spectroscopic redshift ($z_{spec}$) are available.
Using such information, we define the cluster redshift $z_o$
as the mean redshift of the galaxies on the red sequence. 
In Tables \ref{tb:bcg1} and \ref{tb:bcg2} we record the cluster redshifts thus determined, and plot the comparison of $z_o$ with $z_p$ in Figs.~\ref{fig:4} and \ref{fig:5}.
For the low-$z$ bin the mean difference between $z_o$ and $z_p$ is $0.007$, 
while for the high-$z$ bin the mean difference is $0.017$. 
Although the difference between $z_o$ and $z_p$ is small, as a check for systematics,
in the following we will present results using the two estimates separately.

\begin{landscape}
\begin{table*}
\begin{center}
\caption{
Derived properties of clusters and BCGs in the low-$z$ bin.
Columns 1--2 are ID and redshift in {\it Planck} SZ catalogue. 
Column 3 is our estimation of cluster redshift, using red sequence galaxies. 
Column 4 is objID of the identified BCGs in the SDSS database.
Columns 5--12 are estimated stellar masses of the BCGs in unit of $10^{11}h^{-1}M_{\odot}$. 
For the notation $M_{xyz}$,
the 1st subscript denotes the code used for the estimation of stellar mass ($h$ and $k$ stand for {\tt NewHyperz} and {\tt kcorrect}, respectively). 
The 2nd subscript represents the magnitude used ($p$ and $t$ are Petrosian and total magnitude, respectively). 
The 3rd subscript denotes the redshift used ($p$ and $o$ are $z_{p}$ and $z_{o}$, respectively). 
For example, $M_{hpp}$ means the mass obtained by {\tt NewHyperz} using Petrosian magnitude and $z_p$. }
\label{tb:bcg1}
\begin{tabular}{ccccrrrrrrrrc}
\hline\hline
Planck ID & $z_p$ & $z_o$ & objid & $M_{hpp}$ & $M_{htp}$ & $M_{hpo}$ & $M_{hto}$ 
& $M_{kpp}$ & $M_{ktp}$ & $M_{kpo}$ & $M_{kto}$ & spec-$z$\\ 
\hline    
 153    & 0.164    & 0.156    & 1237655473438720267    &   4.05 &   4.25 &   4.05 &   4.25 &      3.36 &   3.53 &   2.96 &   3.11 & 0.16   \\ 
 174    & 0.224    & 0.217    & 1237665569301987827    &  17.97 &  18.50 &  17.16 &  17.66 &      6.96 &   7.16 &   6.47 &   6.66 & 0.223   \\ 
 177    & 0.231    & 0.216    & 1237678596464050454    &   4.12 &   5.36 &   4.41 &   5.75 &      1.63 &   2.12 &   1.41 &   1.84 & 0.231   \\ 
 224    & 0.171    & 0.164    & 1237662194537201844    &   7.91 &  10.00 &   7.91 &  10.00 &      3.02 &   3.81 &   2.34 &   2.96 & 0.17   \\ 
 242    & 0.228    & 0.219    & 1237651715872325879    &   5.06 &   5.17 &   4.83 &   4.94 &      2.23 &   2.28 &   1.72 &   1.76 & $\cdots$   \\ 
 243    & 0.143    & 0.137    & 1237678536348991909    &   5.27 &   5.58 &   5.15 &   5.45 &      1.99 &   2.10 &   1.82 &   1.93 & $\cdots$   \\ 
 248    & 0.232    & 0.231    & 1237680297268019748    &   2.74 &   2.87 &   2.74 &   2.87 &      2.08 &   2.18 &   2.04 &   2.14 & 0.23   \\ 
 256    & 0.147    & 0.150    & 1237678580353531955    &   3.35 &   3.53 &   4.12 &   4.34 &      1.21 &   1.27 &   1.26 &   1.33 & $\cdots$   \\ 
 277    & 0.183    & 0.178    & 1237665328782901386    &   5.48 &   7.45 &   5.48 &   7.45 &      2.02 &   2.75 &   1.91 &   2.59 & $\cdots$   \\ 
 319    & 0.227    & 0.220    & 1237662306722447498    &  10.60 &  11.96 &  10.60 &  11.96 &      3.19 &   3.61 &   2.97 &   3.35 & 0.228   \\ 
 422    & 0.225    & 0.215    & 1237661361296310423    &   9.34 &   9.74 &   9.34 &   9.74 &      3.64 &   3.80 &   3.29 &   3.43 & 0.218   \\ 
 454    & 0.197    & 0.189    & 1237678789202084148    &   5.38 &   5.48 &   5.02 &   5.12 &      1.47 &   1.50 &   1.35 &   1.37 & $\cdots$   \\ 
 530    & 0.136    & 0.134    & 1237657222560874676    &   7.34 &   9.19 &   7.34 &   9.19 &      2.67 &   3.34 &   2.56 &   3.20 & 0.135   \\ 
 531    & 0.169    & 0.161    & 1237678583579279462    &   6.09 &   6.23 &   6.09 &   6.23 &      1.95 &   1.99 &   1.74 &   1.78 & $\cdots$   \\ 
 533    & 0.181    & 0.183    & 1237666464267894977    &   5.12 &   5.92 &   5.12 &   5.92 &      1.86 &   2.16 &   1.91 &   2.21 & $\cdots$   \\ 
 567    & 0.158    & 0.147    & 1237657589241610352    &   9.63 &  10.19 &   9.63 &  10.19 &      3.12 &   3.31 &   2.67 &   2.83 & $\cdots$   \\ 
 572    & 0.144    & 0.140    & 1237657857139999015    &   5.27 &   5.36 &   5.27 &   5.36 &      2.71 &   2.76 &   2.06 &   2.10 & 0.142   \\ 
 578    & 0.217    & 0.211    & 1237655109440241892    &  12.08 &  13.43 &  12.08 &  13.43 &      4.57 &   5.08 &   4.31 &   4.80 & $\cdots$   \\ 
 610    & 0.213    & 0.207    & 1237665126931234947    &   6.77 &   7.10 &   7.09 &   7.44 &      3.23 &   3.39 &   2.53 &   2.65 & 0.214   \\ 
 617    & 0.206    & 0.199    & 1237661139034046481    &   9.67 &   9.97 &   9.67 &   9.97 &      3.30 &   3.41 &   3.07 &   3.17 & 0.206   \\ 
 718    & 0.135    & 0.133    & 1237668288540639614    &   2.29 &   2.82 &   2.29 &   2.82 &      0.96 &   1.19 &   0.93 &   1.14 & 0.136   \\ 
 726    & 0.175    & 0.160    & 1237671260126576915    &   8.45 &   8.77 &   7.03 &   7.29 &      2.54 &   2.63 &   2.09 &   2.17 & 0.176   \\ 
 758    & 0.142    & 0.140    & 1237667783906033793    &  11.56 &  12.20 &  11.56 &  12.20 &      4.04 &   4.26 &   3.93 &   4.14 & $\cdots$   \\ 
 951    & 0.133    & 0.134    & 1237651755084087489    &  13.07 &  14.72 &  13.07 &  14.72 &      3.78 &   4.26 &   3.79 &   4.27 & $\cdots$   \\ 
 988    & 0.165    & 0.156    & 1237658493356408883    &  10.93 &  11.14 &   9.09 &   9.27 &      3.23 &   3.29 &   2.90 &   2.96 & 0.17   \\ 
1105    & 0.183    & 0.174    & 1237655500272500810    &   5.64 &   5.75 &   5.64 &   5.75 &      2.59 &   2.65 &   2.87 &   2.93 & $\cdots$   \\ 
1130    & 0.259    & 0.232    & 1237651736303370409    &   9.80 &  10.59 &   8.53 &   9.23 &      3.26 &   3.53 &   2.52 &   2.73 & 0.26   \\ 
1182    & 0.252    & 0.239    & 1237651754560454806    &   4.23 &   4.51 &   4.04 &   4.30 &      2.47 &   2.64 &   2.14 &   2.28 & 0.252   \\ 
1216    & 0.215    & 0.216    & 1237655497600467190    &   1.68 &   1.73 &   1.68 &   1.73 &      3.38 &   3.48 &   3.43 &   3.53 & 0.217   \\ 
1227    & 0.171    & 0.164    & 1237667781231706367    &   3.99 &   4.61 &   3.32 &   3.84 &      1.20 &   1.39 &   1.10 &   1.27 & 0.173   \\ 
\hline
\end{tabular}
\end{center}
\end{table*} 
\end{landscape}

\begin{landscape}
\begin{table*}
\begin{center}
\caption{
Derived properties of clusters and BCGs in the high-$z$ bin.
Columns 1--2 are ID and redshift in {\it Planck} SZ catalogue. 
Column 3 is our estimation of cluster redshift, using red sequence galaxies. 
Column 4 is objID of the identified BCGs in the SDSS database.
Columns 5--12 are estimated stellar masses of the BCGs in unit of $10^{11}h^{-1}M_{\odot}$. 
For the notation $M_{xyz}$,
the 1st subscript denotes the code used for the estimation of stellar mass ($h$ and $k$ stand for {\tt NewHyperz} and {\tt kcorrect}, respectively). 
The 2nd subscript represents the magnitude used ($p$ and $t$ are Petrosian and total magnitude, respectively). 
The 3rd subscript denotes the redshift used ($p$ and $o$ are $z_{p}$ and $z_{o}$, respectively). 
For example, $M_{kto}$ means the mass obtained by {\tt kcorrect} using total magnitude and $z_o$. }
\label{tb:bcg2}
\begin{tabular}{ccccrrrrrrrrc}
\hline\hline
Planck ID & $z_p$ & $z_o$ & objid  & $M_{hpp}$ & $M_{htp}$  & $M_{hpo}$ & $M_{hto}$ 
 & $M_{kpp}$ & $M_{ktp}$  & $M_{kpo}$ & $M_{kto}$ & spec-$z$\\ 
\hline 
  12    & 0.380    & 0.370    & 1237662264319738193    &  16.68 &  17.09 &  18.29 &  18.74 &      2.69 &   2.75 &   2.53 &   2.60 & 0.375   \\ 
  57    & 0.404    & 0.403    & 1237668657908878169    &   2.61 &   2.64 &   2.61 &   2.64 &      1.33 &   1.34 &   1.31 &   1.32 & $\cdots$   \\ 
  97    & 0.361    & 0.351    & 1237662698117726968    &   5.65 &   5.89 &   5.65 &   5.89 &      2.59 &   2.70 &   2.05 &   2.13 & $\cdots$   \\ 
 128    & 0.427    & 0.386    & 1237662335184667152    &   6.83 &  12.61 &   6.67 &  12.33 &      2.75 &   5.08 &   2.18 &   4.03 & 0.427   \\ 
 137    & 0.389    & 0.376    & 1237651250453349766    &  10.37 &  10.76 &  10.61 &  11.01 &      4.00 &   4.15 &   3.69 &   3.83 & $\cdots$   \\ 
 172    & 0.391    & 0.382    & 1237678618474972017    &   9.79 &  10.33 &   8.93 &   9.42 &      3.81 &   4.01 &   3.60 &   3.79 & $\cdots$   \\ 
 178    & 0.426    & 0.397    & 1237665584873930855    &   6.22 &   6.68 &   5.68 &   6.09 &      3.19 &   3.42 &   2.31 &   2.48 & $\cdots$   \\ 
 181    & 0.447    & 0.391    & 1237652599036838395    &   3.46 &   3.85 &   3.23 &   3.59 &      1.39 &   1.54 &   1.02 &   1.14 & $\cdots$   \\ 
 183    & 0.397    & 0.391    & 1237680066416148649    &  14.85 &  15.61 &  14.85 &  15.61 &      6.17 &   6.49 &   5.99 &   6.30 & $\cdots$   \\ 
 196    & 0.387    & 0.378    & 1237662701872612077    &   4.45 &   4.60 &   4.45 &   4.60 &      2.52 &   2.60 &   2.41 &   2.50 & $\cdots$   \\ 
 234    & 0.400    & 0.393    & 1237659324952216190    &   2.49 &   5.18 &   2.49 &   5.18 &      0.95 &   1.97 &   0.91 &   1.90 & $\cdots$   \\ 
 260    & 0.409    & 0.393    & 1237668610660172385    &   3.87 &   4.35 &   3.45 &   3.88 &      2.06 &   2.32 &   1.57 &   1.76 & $\cdots$   \\ 
 273    & 0.393    & 0.384    & 1237652901820957259    &   3.62 &   3.67 &   3.31 &   3.34 &      1.67 &   1.69 &   1.59 &   1.61 & 0.394   \\ 
 275    & 0.412    & 0.374    & 1237680298882433199    &   7.43 &   7.52 &   9.14 &   9.25 &      3.16 &   3.20 &   2.49 &   2.52 & $\cdots$   \\ 
 324    & 0.352    & 0.375    & 1237672795040645348    &   5.24 &   5.32 &   5.24 &   5.32 &      2.31 &   2.35 &   2.68 &   2.72 & $\cdots$   \\ 
 366    & 0.360    & 0.380    & 1237678596480762080    &   6.90 &   7.34 &   7.06 &   7.51 &      2.77 &   2.95 &   3.16 &   3.36 & 0.365   \\ 
 392    & 0.415    & 0.384    & 1237679476396655293    &   2.40 &   2.42 &   2.24 &   2.26 &      1.12 &   1.13 &   0.95 &   0.96 & $\cdots$   \\ 
 418    & 0.350    & 0.358    & 1237678433792295224    &   3.77 &   3.86 &   3.77 &   3.86 &      1.92 &   1.96 &   2.06 &   2.10 & $\cdots$   \\ 
 427    & 0.389    & 0.384    & 1237678602382344780    &   3.06 &   3.24 &   2.99 &   3.16 &      1.81 &   1.92 &   1.75 &   1.85 & $\cdots$   \\ 
 469    & 0.423    & 0.400    & 1237666216227963460    &   7.61 &   7.83 &   6.78 &   6.98 &      2.64 &   2.72 &   2.34 &   2.41 & $\cdots$   \\ 
 581    & 0.405    & 0.377    & 1237670956790644941    &  11.16 &  11.55 &   9.94 &  10.29 &      4.82 &   4.98 &   4.98 &   5.15 & 0.405   \\ 
 596    & 0.373    & 0.389    & 1237678890137747779    &   6.30 &   6.61 &   6.30 &   6.61 &      2.46 &   2.59 &   3.27 &   3.43 & 0.373   \\ 
 613    & 0.393    & 0.380    & 1237660241387454756    &   4.06 &   6.21 &   7.92 &  12.11 &      2.34 &   2.41 &   2.19 &   2.25 & 0.384   \\ 
 631    & 0.378    & 0.372    & 1237658192150987164    &   4.76 &   5.72 &   4.65 &   5.59 &      2.44 &   2.93 &   2.34 &   2.81 & 0.376   \\ 
 637    & 0.397    & 0.388    & 1237654390561112852    &   4.75 &   4.37 &   7.19 &   6.61 &      2.71 &   2.96 &   2.57 &   2.81 & $\cdots$   \\ 
 642    & 0.380    & 0.370    & 1237665017385910485    &   7.32 &   7.66 &   6.99 &   7.31 &      3.80 &   3.97 &   3.05 &   3.19 & 0.381   \\ 
 674    & 0.355    & 0.363    & 1237673807040022256    &   6.29 &   6.95 &   6.01 &   6.64 &      2.35 &   2.60 &   2.94 &   3.25 & $\cdots$   \\ 
 715    & 0.382    & 0.370    & 1237667733956395341    &   7.45 &   8.21 &   7.45 &   8.21 &      3.08 &   3.39 &   2.86 &   3.15 & 0.383   \\ 
 888    & 0.412    & 0.386    & 1237667783373095073    &   3.85 &   3.87 &   3.77 &   3.79 &      2.16 &   2.17 &   1.96 &   1.98 & $\cdots$   \\ 
1120    & 0.425    & 0.391    & 1237662239082349083    &   5.72 &   5.87 &   6.28 &   6.43 &      3.15 &   3.23 &   2.23 &   2.29 & 0.426   \\ 
\hline
\end{tabular}
\end{center}
\end{table*} 
\end{landscape}

Although BCGs are typically red and dead, and thus belonging to the red sequence, in some cases they appear blue, likely due to star formation from cooling gas \citep[e.g.,][]{fabian94,odea08}. 
To account for this, 
when we look for the BCGs, in addition to using the available $z_{phot}$ and $z_{spec}$ information,
 we also consider a 
wider colour range, $-1.0< (g-i)-(g-i)_{\rm fit}<1.0$.
As the final step, we visually inspect the colour images of BCG candidates and manually correct for misidentified BCGs.  We give preference to galaxies with early type morphology and are closer to the centre of galaxy concentration.
The parameters of the identified BCGs are listed in Tables \ref{tb:bcg1} and \ref{tb:bcg2}.

In Section \ref{assumption} we have defined the two redshift bins that will be used to study the BCG evolution.
Within the SDSS DR10 footprint, there are 121 and 30 clusters in the low-$z$ and high-$z$ bins, respectively, for which we can find an optical counterpart.  (The numbers of clusters remain the same no matter whether $z_p$ or $z_o$ is used.)
Given the small number of clusters in the high-$z$ bin, we therefore focus on top $N=30$ clusters in both bins.
For the low-$z$ bin, we sort the clusters by the mass proxy $M_{Y_Z}$, and select the most massive 30 clusters as our sample.
The clusters in the high-$z$ (low-$z$) bin have $M_{Y_Z}\ge 3.6\times 10^{14} M_\odot$ ($2.2\times 10^{14} M_\odot$).
Given the overlap between SDSS DR10 and PSZ catalog, which we roughly estimate to be $\sim 13000$ deg$^2$, the comoving volume of each redshift bin is about $(770 h^{-1}$\,Mpc$)^3$.

\subsection{Estimation of stellar mass of BCGs}
\label{bcg_sm}

We estimate the stellar mass of BCGs from SDSS data by two spectral energy distribution (SED) fitting techniques.  
The first tool we use is the {\tt kcorrect} package \citep[v4.2;][]{2007AJ....133..734B}, while the second one is the code  
{\tt NewHyperz} (v11)\footnote{http://www.ast.obs-mip.fr/users/roser/hyperz/}. 
Having two independent methods allows us to assess the robustness of our results, as well as to evaluate any systematic uncertainty in the stellar mass estimates, as such estimates inevitably depend on the chosen templates (or/and stellar population synthesis libraries) and the initial mass function (IMF), among other factors.

For {\tt kcorrect},
we use both the default and the luminous red galaxy templates, both of which are constructed by \citet{2007AJ....133..734B} using the \citet[][hereafter BC03]{2003MNRAS.344.1000B} models with the \citet{2003PASP..115..763C} IMF. 
The default templates are constructed from a set of 485 BC03 models that span a wide range of star formation histories and metallicities.
The nonnegative linear combinations of these and the LRG templates are shown to be able to describe the great majority of observed galaxy spectra from SDSS and other surveys \citep{2007AJ....133..734B}.
We use extinction corrected $ugriz$ model magnitudes and either $z_o$ or $z_p$ when fitting the BCG SED. 
As a sanity check,
we compare our derived stellar masses from {\tt kcorrect} with those estimated by  \citet{2012MNRAS.421..314C}, 
who have developed a variant of the principle component analysis of spectral decomposition, so that they are able to infer physical quantities (e.g., stellar mass, metallicity) from the eigenspectra for each of the galaxies.
Their stellar mass estimates are
available in the table {\tt stellarMassPCAWiscBC03} in SDSS DR10. 
Among the 60 BCGs in our sample, 27 of them have entries in this table. 
Although there is a small offset between these two sets of stellar masses, with the DR10 stellar masses being slightly higher than the {\tt kcorrect} ones, 
the scatter is small and the correlation is clear.  

For {\tt NewHyperz}, 
we use the BC03 early-type templates in the SED fitting.
We assume the BCG SEDs are described by a single stellar population, and thus we do not
fit multiple templates simultaneously.
Again the extinction corrected $ugriz$ model magnitudes are used for the fitting.
When the template fitting finishes, {\tt NewHyperz} provides a scaling factor for the best-fit SED. Since the BC03 templates are in units of solar luminosities per solar mass, the stellar mass can be easily derived from the scaling factor.

Although the SDSS model magnitudes are suitable for SED fitting (as they are calculated within the same circular aperture, defined in $r$-band), they may not capture the total light from the galaxies.
Petrosian magnitudes, on the other hand, are designed to capture the same fraction of the light irrespective of the distance to the galaxies, so they provide a good way to compare galaxy stellar masses at different redshifts (provided that the galaxies have similar profiles).  Ideally, one wants to compare the ``total'' luminosity or stellar mass of the galaxies.  
We follow the method of \citet{2005AJ....130.1535G} to extrapolate from Petrosian magnitudes to total magnitudes.
Basically, from the observed radii $r_{50}$ and $r_{90}$, which enclose 50\% and 90\% of the Petrosian flux, one can infer the Sersic index $n$, which then enables one to calculate the magnitude difference between the Petrosian magnitude and the ``total'' magnitude (assuming the galaxy surface brightness distribution follows the Sersic profile).

Once stellar mass based on model magnitudes ($M_{*m}$) is derived, we can further convert it to those based on Petrosian and total magnitudes ($M_{*p}$ and $M_{*t}$), from the magnitude differences.  That is,
\begin{equation}
M_{*x}=10^{-0.4(m_x-m_m)}M_{*m}, 
\label{eq:4}
\end{equation}
where $m_m$ is the apparent model magnitude, $m_x$ is either Petrosian or total magnitude ($x=p$ or $t$).

\section{The Results}
\label{results_1}

To accommodate various modeling and observational uncertainties, we have decided to present results considering both the {\it Planck}-based and our own cluster redshift estimates, and to perform SED fitting with two independent codes ({\tt kcorrect} and {\tt NewHyperz}), as well as inferring stellar masses from the Petrosian or total magnitudes.

The stellar mass estimates from the combination of all these choices are presented in Tables 1 and 2, for the sake of completeness.

\begin{figure*}
\includegraphics[width=.65\textwidth]{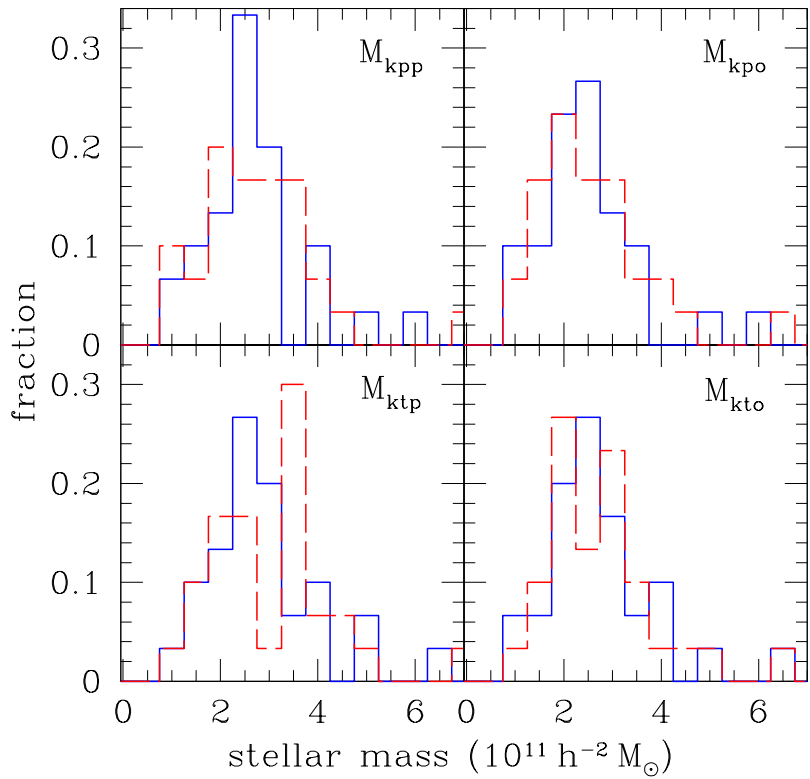}
\caption{
The distribution of the BCG stellar mass estimated using {\tt kcorrect}.
The solid and dashed lines are for the high-$z$ and low-$z$ bins, respectively.   
Panels in the top and bottom rows are for the mass estimation based on Petrosian and total magnitudes, respectively, while 
stellar mass in the left and right panels is estimated by using $z_p$ and $z_o$, respectively.
The label to the upper right corner of each panel is the same as that in Tables 1 \& 2.
}
\label{fig:k-correct}
\end{figure*}

We present the BCG stellar mass estimates in both redshift bins using {\tt kcorrect} in Fig. \ref{fig:k-correct}, and tabulate the resulting differences in the mean stellar masses between two redshift bins in Table \ref{tb:result3}.  As can be seen in the Table, our results do not depend sensitively on the way cluster redshifts are estimated ($z_p$ vs $z_o$), and the {\it mean} stellar mass growth is of order a few percent irrespective of the magnitude measurement (Petrosian or total) from which it is derived.
We note that, for all the values shown in Tables \ref{tb:result3}, they are consistent with zero.

\begin{figure*}
\includegraphics[width=.65\textwidth]{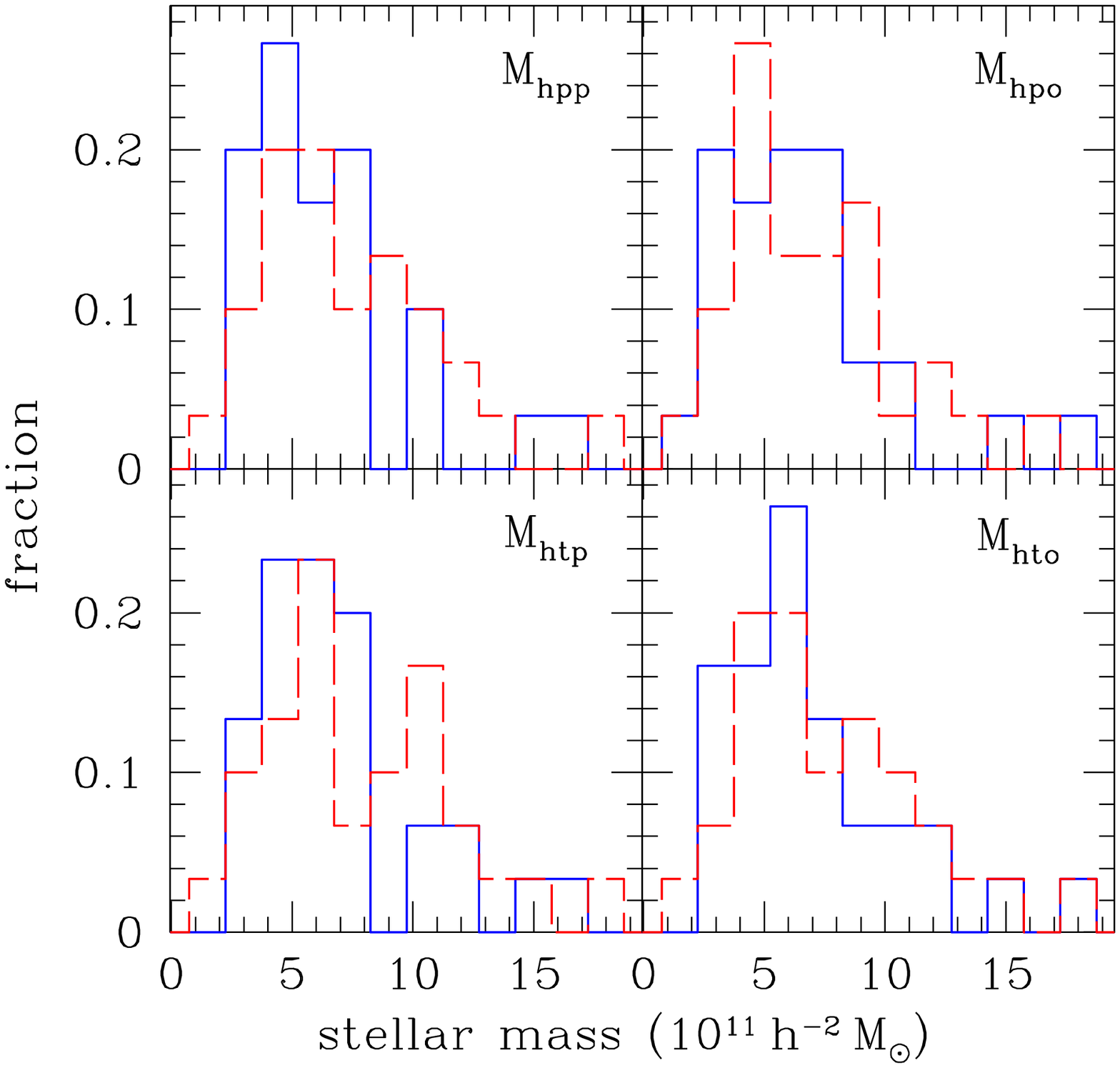}
\caption{
The distribution of the BCG stellar mass estimated using {\tt NewHyperz}.
The solid and dashed lines are for the high-$z$ and low-$z$ bins, respectively.   
Panels in the top and bottom rows are for the mass estimation based on Petrosian and total magnitudes, respectively, while 
stellar mass in the left and right panels is estimated by using $z_p$ and $z_o$, respectively.
The label to the upper right corner of each panel is the same as that in Tables 1 \& 2.
}
\label{fig:NewHyperz}
\end{figure*}

The results based on {\tt NewHyperz} are shown in Fig.~\ref{fig:NewHyperz} and also in Table \ref{tb:result3}.  
Comparing to the results from {\tt correct}, the {\tt NewHyperz}-based values are larger, which may be due to the use of just the early-type template.

As mentioned above, Petrosian magnitudes provide a meaningful way to compare galaxies at different redshifts; considering all possible combinations of ways to estimate stellar mass using Petrosian magnitudes, we conclude that the observationally determined stellar mass growth between $z\sim 0.4$ to $z\sim 0.2$ is $2-14\%$, and very likely in the lower part of this range, 
once we consider the small bias suggested by the simulation presented in Section \ref{bcg_growth}.

In addition to comparing the mean of the distributions of BCG masses, we also use the Kolmogorov-Smirnov (KS) test to examine whether the BCG mass distributions at $z=0.4$ and $z=0.2$ are similar.  For {\tt kcorrect}-based stellar masses, the KS test indicates the two distributions are the same at $\sim 96\%$, regardless of the magnitudes used.  For {\tt NewHyperz}-based masses, although the likelihood is lower ($\sim 35\%$), the two distributions are still quite possible to be drawn from the same parent distribution.

\begin{table*}
\begin{center}
\caption{The mean BCG stellar mass growth between $z=0.2$ and $z=0.4$ obtained by {\tt kcorrect} (2nd and 3rd columns) and by {\tt NewHyperz} (4th and 5th columns).
The uncertainties quoted are statistical only, and are obtained from the mean absolute error. 
 Top and bottom rows show the mass growth based on Petrosian and total magnitudes, respectively.}
\begin{tabular}{c|cc|cc}
\hline\hline
 & \multicolumn{2}{c}{{\tt kcorrect}} & \multicolumn{2}{c}{{\tt NewHyperz}} \\
\hline
  & mean($z=z_{p}$) & mean($z=z_{o}$) & mean($z=z_{p}$) & mean($z=z_{o}$) \\
\hline
 Petrosian-mag & 4.1\%$\pm$8.3\%  & 1.5\%$\pm$8.4\%  &  13.7\%$\pm$12.3\% & 7.9\%$\pm$11.4\% \\
\hline
 total-mag & 3.7\%$\pm$8.2\%  & 1.2\%$\pm$8.3\%  & 12.6\%$\pm$12.1\% &  6.6\%$\pm$11.5\%\\
\hline\hline
\end{tabular}
\label{tb:result3}
\end{center}
\end{table*}

\begin{figure}
\includegraphics[width=.45\textwidth]{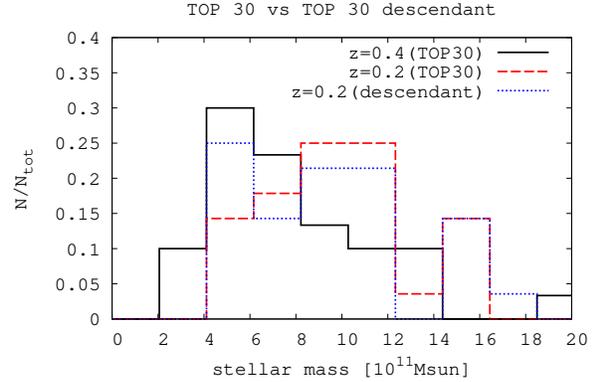}
\caption{The mass distribution of BCGs in the model of \citet{2011MNRAS.413..101G}. 
The black solid histogram is for top 30 most massive halos at $z=0.4$. 
The red dashed histogram is for the top 30 most massive halos at $z=0.2$.
The blue dotted histogram is for the descendants of top 30 $z=0.4$ halos. }
\label{fig:15}
\end{figure}

Finally, we compare these results with those obtained from the Millennium Run simulation.  
In particular, we use the \citet{2011MNRAS.413..101G} model to examine the evolution of model BCGs from $z=0.4$ to $z=0.2$.
The central galaxies in massive halos are regarded as the BCGs.
Similar to the test presented in Section \ref{bcg_growth},
there are two ways to select the model clusters. 
The first one is to select top $N$ massive clusters at each redshift. 
The other way is to follow the merger trees and study the descendant halos of top $N$ $z=0.4$ halos. 
For the first method, the mean mass of the BCGs of top 30 halos  
increases by $38\%$ from $z=0.4$ to $z=0.2$ (Fig.~\ref{fig:15}, red dashed histogram vs black solid histogram). 
On the other hand, for the second method, 
the mean mass of the BCGs  
increases by $31\%$ (Fig.~\ref{fig:15}, blue dotted histogram vs black solid histogram). 
We can thus conclude that in the  \citet{2011MNRAS.413..101G} model the typical stellar mass growth of BCGs is of order 30\%, which is quite large compared to the observed value (likely much less than $14\%$).
This finding is consistent with the results of \citet{2013ApJ...771...61L} and \citet{2012MNRAS.427..550L}, and indicates that the degree of late time ($z<0.5$) growth is small.

\section{Summary}
\label{summary}

In this paper, we have studied the stellar mass growth of 
BCGs in massive clusters at $z<0.5$. 
Although it is impossible to follow the growth of any single galaxy,
we have developed a statistical approach that allows us to construct samples of BCGs that can be regarded as progenitor-descendant pairs.
This is based on the Ansatz
that the top $N$ most massive clusters at one redshift largely remain in top $N$ most massive clusters at a slightly later cosmic epoch (Section \ref{assumption}).
If this Ansatz holds, the top $N$ most massive clusters observed at one redshift
can be regarded as representative of the progenitors of top $N$ most massive clusters found at a lower redshift.
Our simulations suggest that between $z=0.2$ and $z=0.4$, with $N=30$, this Ansatz holds to about $75\%$ for comoving volume of $\gtrsim (700 h^{-1}$\,Mpc$)^3$.
Using simulations, we have found that the mean growth of BCGs inferred this way may be slightly biased high (by $\sim 5-10\%$ or so) compared to the actual growth (Section \ref{bcg_growth}).

This idea has been applied to {\it Planck} clusters that lie within SDSS DR10 footprint.  We consider two redshift bins ($0.13\le z \le0.26$ and $0.37 \le z \le 0.41$) that occupy the same comoving volume, and select top 30 most massive clusters in each bin.
We have identified the BCGs in these clusters by considering their location on the colour-magnitude diagram, their morphology, and their proximity to the cluster centre.
The stellar mass of the BCGs is estimated by two different SED fitting codes, {\tt kcorrect} and {\tt NewHyperz}.
In addition to the way SED fitting is done, we also consider various ways the luminosity is measured (Petrosian mag and total mag), which affect the resulting stellar mass estimates.
Considering all these choices and their observational consequences, we conclude that the probable stellar mass growth of BCGs in massive clusters from $z\sim 0.4$ to $z\sim 0.2$ is $2-14\%$, and very likely to be just few percent. 
We emphasise that 
although the two SED fitting codes do not give the same masses, what is more relevant is the relative stellar mass growth between $z=0.4$ and $z=0.2$.

We compare the observational results with the predictions from a
SAM built upon the Millennium Run simulation \citep{2011MNRAS.413..101G}. 
Considering different ways we can associate halos at one
redshift with those in another, we conclude the model BCGs typically
have grown by about 30\% from $z \sim 0.4$ to $z \sim 0.2$.
We note that there is no spatial information in the model of \citet{2011MNRAS.413..101G}, and the galaxy growth is calculated by
accretion of satellite galaxies at all radii.
Observationally, we have attempted to compute the ``total'' stellar
mass growth in an approximate way via the total magnitude, and found
that at most the growth is about 13\%. This is based on extrapolation
of the photometry, and may depend on the accuracy of sky subtraction
in crowded fields such as cluster cores.
It is possible that our estimate of growth is still confined to
regions that are relatively small compared to the true total size that
is considered in SAMs, especially for intrinsically large galaxies such
as BCGs.

In summary, the lack of spatial information in current SAMs makes it
non-trivial to compare with our data.  Perhaps a better way to proceed
is to compare with hydrodynamical simulations,  or simulations with
the ``tagging'' technique \citep{cooper14}, which would provide spatial density
profiles of the model galaxies, making a fairer comparison possible.

Our Ansatz has enabled us to ``link'' the most massive clusters at different redshifts together.  
We plan to apply this methodology to the on-going Subaru HyperSuprimeCam survey 
\citep{2010AIPC.1279..120T}, which will image 1400 deg$^2$ of the sky to $r\sim 26$, and will provide a high quality cluster sample out to $z\sim 1.5$.  We will then be able to trace the way massive galaxies evolve in the most massive, densest environments in the past 9 Gyr.

\section*{Acknowledgements}

We are grateful to Jerry Ostriker for suggesting the Ansatz used in this work, and to an anonymous referee for comments that improved the presentation of the paper.
This work is supported in part by JSPS
Grant-in-Aid for the Global COE programs, ``Quest for
Fundamental Principles in the Universe: from Particles to
the Solar System and the Cosmos'' at Nagoya University.
Y.-T.~L. acknowledges support from the Ministry of Science and Technology grant NSC 102-2112-M-001-001-MY3.
NS is supported by Grand-in-Aid for Scientific Research No.~22340056
and 18072004.
Funding for SDSS-III has been provided by the Alfred P.~Sloan Foundation, the Participating Institutions, the National Science Foundation, and the U.S.~Department of Energy Office of Science. The SDSS-III web site is http://www.sdss3.org/.
SDSS-III is managed by the Astrophysical Research Consortium for the Participating Institutions of the SDSS-III Collaboration.
The Millennium Simulation databases used in this paper and the web application providing online access to them were constructed as part of the activities of the German Astrophysical Virtual Observatory (GAVO).

\bibliographystyle{mn2e}  

\begin{thebibliography}{}

\bibitem[\protect\citeauthoryear{{Ahn}, {Alexandroff}, {Allende Prieto},
  {Anders}, {Anderson}, {Anderton}, {Andrews}, {Aubourg}, {Bailey}, {Bastien}
  \& et al.}{{Ahn} et~al.}{2013}]{2013arXiv1307.7735A}
{Ahn} C.~P.,  {Alexandroff} R.,  {Allende Prieto} C.,  {Anders} F.,  {Anderson}
  S.~F.,  {Anderton} T.,  {Andrews} B.~H.,  {Aubourg} {\'E}.,  {Bailey} S.,
  {Bastien} F.~A.,    et al. 2013, ArXiv e-prints


\bibitem[\protect\citeauthoryear{{Blanton} \& {Roweis}}{{Blanton} \&
  {Roweis}}{2007}]{2007AJ....133..734B}
{Blanton} M.~R.,  {Roweis} S.,  2007, \aj, 133, 734

\bibitem[\protect\citeauthoryear{{Bruzual} \& {Charlot}}{{Bruzual} \&
  {Charlot}}{2003}]{2003MNRAS.344.1000B}
{Bruzual} G.,  {Charlot} S.,  2003, \mnras, 344, 1000

\bibitem[\protect\citeauthoryear{{Chabrier}}{{Chabrier}}{2003}]{2003PASP..115..763C}
{Chabrier} G.,  2003, \pasp, 115, 763

\bibitem[\protect\citeauthoryear{{Chen}, {Kauffmann}, {Tremonti}, {White},
  {Heckman}, {Kova{\v c}}, {Bundy}, {Chisholm}, {Maraston}, {Schneider},
  {Bolton}, {Weaver} \& {Brinkmann}}{{Chen} et~al.}{2012}]{2012MNRAS.421..314C}
{Chen} Y.-M.,  {Kauffmann} G.,  {Tremonti} C.~A.,  {White} S.,  {Heckman}
  T.~M.,  {Kova{\v c}} K.,  {Bundy} K.,  {Chisholm} J.,  {Maraston} C.,
  {Schneider} D.~P.,  {Bolton} A.~S.,  {Weaver} B.~A.,    {Brinkmann} J.,
  2012, \mnras, 421, 314

\bibitem[\protect\citeauthoryear{{Collins}, {Stott}, {Hilton}, {Kay},
  {Stanford}, {Davidson}, {Hosmer}, {Hoyle}, {Liddle}, {Lloyd-Davies}, {Mann},
  {Mehrtens}, {Miller}, {Nichol}, {Romer}, {Sahl{\'e}n}, {Viana} \&
  {West}}{{Collins} et~al.}{2009}]{2009Natur.458..603C}
{Collins} C.~A.,  {Stott} J.~P.,  {Hilton} M.,  {Kay} S.~T.,  {Stanford} S.~A.,
   {Davidson} M.,  {Hosmer} M.,  {Hoyle} B.,  {Liddle} A.,  {Lloyd-Davies} E.,
  {Mann} R.~G.,  {Mehrtens} N.,  {Miller} C.~J.,  {Nichol} R.~C.,  {Romer}
  A.~K.,  {Sahl{\'e}n} M.,  {Viana} P.~T.~P.,    {West} M.~J.,  2009, \nat,
  458, 603


\bibitem[Cooper et al.(2014)]{cooper14} Cooper, A.~P., Gao, L., 
Guo, Q., et al.\ 2014, \mnras, submitted (arXiv:1407.5627) 

\bibitem[Davis et al.(1985)]{davis85} Davis, M., Efstathiou, 
G., Frenk, C.~S., \& White, S.~D.~M.\ 1985, \apj, 292, 371 




\bibitem[\protect\citeauthoryear{{De Lucia} \& {Blaizot}}{{De Lucia} \&
  {Blaizot}}{2007}]{2007MNRAS.375....2D}
{De Lucia} G.,  {Blaizot} J.,  2007, \mnras, 375, 2

\bibitem[\protect\citeauthoryear{{De Lucia}, {Springel}, {White}, {Croton} \&
  {Kauffmann}}{{De Lucia} et~al.}{2006}]{2006MNRAS.366..499D}
{De Lucia} G.,  {Springel} V.,  {White} S.~D.~M.,  {Croton} D.,    {Kauffmann}
  G.,  2006, \mnras, 366, 499

\bibitem[Fabian(1994)]{fabian94} Fabian, A.~C.\ 1994, ARA\&A, 32, 277 

\bibitem[\protect\citeauthoryear{{Gao}, {Loeb}, {Peebles}, {White} \&
  {Jenkins}}{{Gao} et~al.}{2004}]{2004ApJ...614...17G}
{Gao} L.,  {Loeb} A.,  {Peebles} P.~J.~E.,  {White} S.~D.~M.,    {Jenkins} A.,
  2004, \apj, 614, 17

\bibitem[\protect\citeauthoryear{{Graham}, {Driver}, {Petrosian}, {Conselice},
  {Bershady}, {Crawford} \& {Goto}}{{Graham}
  et~al.}{2005}]{2005AJ....130.1535G}
{Graham} A.~W.,  {Driver} S.~P.,  {Petrosian} V.,  {Conselice} C.~J.,
  {Bershady} M.~A.,  {Crawford} S.~M.,    {Goto} T.,  2005, \aj, 130, 1535

\bibitem[\protect\citeauthoryear{{Guo}, {White}, {Boylan-Kolchin}, {De Lucia},
  {Kauffmann}, {Lemson}, {Li}, {Springel} \& {Weinmann}}{{Guo}
  et~al.}{2011}]{2011MNRAS.413..101G}
{Guo} Q.,  {White} S.,  {Boylan-Kolchin} M.,  {De Lucia} G.,  {Kauffmann} G.,
  {Lemson} G.,  {Li} C.,  {Springel} V.,    {Weinmann} S.,  2011, \mnras, 413,
  101

\bibitem[\protect\citeauthoryear{{Lidman}, {Suherli}, {Muzzin}, {Wilson},
  {Demarco}, {Brough}, {Rettura}, {Cox}, {DeGroot}, {Yee}, {Gilbank},
  {Hoekstra}, {Balogh}, {Ellingson}, {Hicks}, {Nantais}, {Noble}, {Lacy},
  {Surace} \& {Webb}}{{Lidman} et~al.}{2012}]{2012MNRAS.427..550L}
{Lidman} C.,  {Suherli} J.,  {Muzzin} A.,  {Wilson} G.,  {Demarco} R.,
  {Brough} S.,  {Rettura} A.,  {Cox} J.,  {DeGroot} A.,  {Yee} H.~K.~C.,
  {Gilbank} D.,  {Hoekstra} H.,  {Balogh} M.,  {Ellingson} E.,  {Hicks} A.,
  {Nantais} J.,  {Noble} A.,  {Lacy} M.,  {Surace} J.,    {Webb} T.,  2012,
  \mnras, 427, 550

\bibitem[Lin 
\& Mohr(2004)]{lin04} Lin, Y.-T., \& Mohr, J.~J.\ 2004, \apj, 617, 879 

\bibitem[Lin et al.(2013)]{2013ApJ...771...61L} Lin, Y.-T., Brodwin, M., 
Gonzalez, A.~H., et al.\ 2013, \apj, 771, 61 


\bibitem[\protect\citeauthoryear{{L{\'o}pez-Cruz}, {Barkhouse} \&
  {Yee}}{{L{\'o}pez-Cruz} et~al.}{2004}]{2004ApJ...614..679L}
{L{\'o}pez-Cruz} O.,  {Barkhouse} W.~A.,    {Yee} H.~K.~C.,  2004, \apj, 614,
  679

\bibitem[\protect\citeauthoryear{{Muzzin}, {Marchesini}, {Stefanon}, {Franx},
  {McCracken}, {Milvang-Jensen}, {Dunlop}, {Fynbo}, {Brammer}, {Labb{\'e}} \&
  {van Dokkum}}{{Muzzin} et~al.}{2013}]{2013ApJ...777...18M}
{Muzzin} A.,  {Marchesini} D.,  {Stefanon} M.,  {Franx} M.,  {McCracken} H.~J.,
   {Milvang-Jensen} B.,  {Dunlop} J.~S.,  {Fynbo} J.~P.~U.,  {Brammer} G.,
  {Labb{\'e}} I.,    {van Dokkum} P.~G.,  2013, \apj, 777, 18

\bibitem[O'Dea et al.(2008)]{odea08} O'Dea, C.~P., Baum, 
S.~A., Privon, G., et al.\ 2008, \apj, 681, 1035 



\bibitem[Planck Collaboration et al.(2013a)]{2013arXiv1303.5062P} Planck 
Collaboration I, Ade, P.~A.~R., Aghanim, N., et al.\ 2013a, arXiv:1303.5062 

  
\bibitem[Planck Collaboration et al.(2013b)]{2013arXiv1303.5080P} Planck 
Collaboration XX, Ade, P.~A.~R., Aghanim, N., et al.\ 2013b, arXiv:1303.5080 



\bibitem[Planck Collaboration et al.(2013c)]{2013arXiv1303.5089P} Planck 
Collaboration XXIX, Ade, P.~A.~R., Aghanim, N., et al.\ 2013c, arXiv:1303.5089 



\bibitem[\protect\citeauthoryear{{Springel}}{{Springel}}{2005}]{2005MNRAS.364.1105S}
{Springel} V.,  2005, MNRAS, 364, 1105

\bibitem[\protect\citeauthoryear{{Springel}, {White}, {Jenkins}, {Frenk},
  {Yoshida}, {Gao}, {Navarro}, {Thacker}, {Croton}, {Helly}, {Peacock}, {Cole},
  {Thomas}, {Couchman}, {Evrard}, {Colberg} \& {Pearce}}{{Springel}
  et~al.}{2005}]{2005Natur.435..629S}
{Springel} V.,  {White} S.~D.~M.,  {Jenkins} A.,  {Frenk} C.~S.,  {Yoshida} N.,
   {Gao} L.,  {Navarro} J.,  {Thacker} R.,  {Croton} D.,  {Helly} J.,
  {Peacock} J.~A.,  {Cole} S.,  {Thomas} P.,  {Couchman} H.,  {Evrard} A.,
  {Colberg} J.,    {Pearce} F.,  2005, \nat, 435, 629

\bibitem[\protect\citeauthoryear{{Springel}, {Yoshida} \& {White}}{{Springel}
  et~al.}{2001}]{2001NewA....6...79S}
{Springel} V.,  {Yoshida} N.,    {White} S.~D.~M.,  2001, New Astronomy, 6, 79

\bibitem[\protect\citeauthoryear{{Stott}, {Collins}, {Sahl{\'e}n}, {Hilton},
  {Lloyd-Davies}, {Capozzi}, {Hosmer}, {Liddle}, {Mehrtens}, {Miller}, {Romer},
  {Stanford}, {Viana}, {Davidson}, {Hoyle}, {Kay} \& {Nichol}}{{Stott}
  et~al.}{2010}]{2010ApJ...718...23S}
{Stott} J.~P.,  {Collins} C.~A.,  {Sahl{\'e}n} M.,  {Hilton} M.,
  {Lloyd-Davies} E.,  {Capozzi} D.,  {Hosmer} M.,  {Liddle} A.~R.,  {Mehrtens}
  N.,  {Miller} C.~J.,  {Romer} A.~K.,  {Stanford} S.~A.,  {Viana} P.~T.~P.,
  {Davidson} M.,  {Hoyle} B.,  {Kay} S.~T.,    {Nichol} R.~C.,  2010, \apj,
  718, 23

\bibitem[\protect\citeauthoryear{{Sunyaev} \& {Zel'dovich}}{{Sunyaev} \&
  {Zel'€™dovich}}{1970}]{1970Ap&SS...7...20S}
{Sunyaev} R.~A.,  {Zel'dovich} Y.~B.,  1970, \apss, 7, 20

\bibitem[\protect\citeauthoryear{{Takada}}{{Takada}}{2010}]{2010AIPC.1279..120T}
{Takada} M.,  2010, in {Kawai} N.,  {Nagataki} S.,  eds, American Institute of
  Physics Conference Series Vol.~1279 of American Institute of Physics
  Conference Series, {Subaru Hyper Suprime-Cam Project}.
pp 120--127

\bibitem[\protect\citeauthoryear{{van Dokkum}, {Whitaker}, {Brammer}, {Franx},
  {Kriek}, {Labb{\'e}}, {Marchesini}, {Quadri}, {Bezanson}, {Illingworth},
  {Muzzin}, {Rudnick}, {Tal} \& {Wake}}{{van Dokkum}
  et~al.}{2010}]{2010ApJ...709.1018V}
{van Dokkum} P.~G.,  {Whitaker} K.~E.,  {Brammer} G.,  {Franx} M.,  {Kriek} M.,
   {Labb{\'e}} I.,  {Marchesini} D.,  {Quadri} R.,  {Bezanson} R.,
  {Illingworth} G.~D.,  {Muzzin} A.,  {Rudnick} G.,  {Tal} T.,    {Wake} D.,
  2010, \apj, 709, 1018

\bibitem[\protect\citeauthoryear{{Whiley}, {Arag{\'o}n-Salamanca}, {De Lucia},
  {von der Linden}, {Bamford}, {Best}, {Bremer}, {Jablonka}, {Johnson},
  {Milvang-Jensen}, {Noll}, {Poggianti}, {Rudnick}, {Saglia}, {White} \&
  {Zaritsky}}{{Whiley} et~al.}{2008}]{2008MNRAS.387.1253W}
{Whiley} I.~M.,  {Arag{\'o}n-Salamanca} A.,  {De Lucia} G.,  {von der Linden}
  A.,  {Bamford} S.~P.,  {Best} P.,  {Bremer} M.~N.,  {Jablonka} P.,  {Johnson}
  O.,  {Milvang-Jensen} B.,  {Noll} S.,  {Poggianti} B.~M.,  {Rudnick} G.,
  {Saglia} R.,  {White} S.,    {Zaritsky} D.,  2008, \mnras, 387, 1253

\bibitem[White \& Rees(1978)]{white78} White, S.~D.~M., \& Rees, M.~J.\ 1978, \mnras, 183, 341 



\bibitem[\protect\citeauthoryear{{York}, {Adelman}, {Anderson} Jr., {Anderson},
  {Annis}, {Bahcall}, {Bakken}, {Barkhouser}, {Bastian} \& {SDSS
  Collaboration}}{{York} et~al.}{2000}]{2000AJ....120.1579Y}
{York} D.~G.,  {Adelman} J.,  {Anderson} Jr. J.~E.,  {Anderson} S.~F.,  {Annis}
  J.,  {Bahcall} N.~A.,  {Bakken} J.~A.,  {Barkhouser} R.,  {Bastian} S.,
  {SDSS Collaboration} 2000, \aj, 120, 1579

\bibitem[\protect\citeauthoryear{{Zel'dovich} \& {Sunyaev}}{{Zel'€™dovich} \&
  {Sunyaev}}{1969}]{1969Ap&SS...4..301Z}
{Zel'dovich} Y.~B.,  {Sunyaev} R.~A.,  1969, \apss, 4, 301

\end{thebibliography}

\bsp

\label{lastpage}

\end{document}